
\def\newpage{\vfill\eject}
\def\centreline{\centerline}

%
\def\ref{\par \noindent \hangindent=3pc \hangafter=1}
\def\he{$^3$He }
\def\hed{D $+^3$He\a}
\def\xet{$X^0_{2{\rm MAX}}$\a}
\def\xe{$X_{2{\rm MAX}}$\a}

%

%

%

%

\def\double_under#1{\underline{\underline{#1}}}

\def\pmb#1{\setbox0=\hbox{#1}%
\kern-.025em\copy0\kern-\wd0
\kern.05em\copy0\kern-\wd0
\kern-.025em\raise.0433em\box0}
\def\coms{\mag=1200 \hsize=15truecm \hoffset=1.0truecm \vskip 1.2truecm
\voffset=1.6truecm \vsize=21truecm \tabskip=1em plus 2.5em minus 0.5em
\parindent 1.0truecm \parskip=10pt}
\def\a{\ }

%
%
%

%
%
%

\def\ref{\par \noindent
\hangindent=1pc
\hangafter=-1}
\def\spose#1{\hbox to 0pt{#1\hss}}
\def\simlt{\mathrel{\spose{\lower 3pt\hbox{$\mathchar"218$}}
     \raise 2.0pt\hbox{$\mathchar"13C$}}}
\def\simgt{\mathrel{\spose{\lower 3pt\hbox{$\mathchar"218$}}
     \raise 2.0pt\hbox{$\mathchar"13E$}}}
\def\lsim{\rlap{$<$}{\lower 1.0ex\hbox{$\sim$}}}
\def\gsim{\rlap{$>$}{\lower 1.0ex\hbox{$\sim$}}}

\def\ni{\noindent}


%
%
\def\ref{\par \noindent \hangindent=3pc \hangafter=1}

\def\ApJ{ApJ\a}
\def\ApJS{ApJS\a}

\def\MN{MNRAS\a}

\mathchardef\twiddle="2218

\def\multleft#1{\hbox to size{\vbox {\halign {\lft{##}\cr #1}}\hfill}\par}
\def\multright#1{\hbox to size{\vbox {\halign {\rt{##}\cr #1}}\hfill}\par}

\def\<{\thinspace}

%
%
%
%

\coms
 \hfill OSU-TA-12/94

 \bigskip
\bigskip
\centreline{\bf GENERIC EVOLUTION OF DEUTERIUM AND HELIUM-3}
\bigskip
\bigskip
\centreline{GARY STEIGMAN}
\bigskip
\centreline{Departments of Physics and Astronomy}
\centreline{The Ohio State University}
\centreline{174 West 18th Avenue, Columbus, OH 43210, USA}
\centreline{ and }
\bigskip
\centreline{ MONICA TOSI}
\bigskip
\centreline{Osservatorio Astronomico di Bologna,}
\centreline{Via Zamboni 33, I-40126 Bologna, ITALY}
\bigskip

 \bigskip
\centerline{\bf Summary}

\noindent

\noindent The primordial abundances of deuterium and of helium-3, produced
during
big bang nucleosynthesis, depend sensitively on the baryon density.
Thus, the observed abundances of D and \he may provide useful
``baryometers'' provided the evolution from primordial to present (or,
presolar nebula) abundances is understood.  Inevitably, the derivation
of primordial from observed abundances requires the intervention of a
model for galactic evolution and, so, the inferred primordial
abundances are, necessarily, model dependent.  Here, an analytic
framework for the evolution of D and \he  is presented which is
``generic'' in the sense that it should describe the results of any
specific galactic evolution model.  The ``effective \he  survival
fraction'', $\Gamma _3$, is the one free parameter which is model specific.
Solar system and interstellar data are used to infer upper and lower
bounds to the primordial deuterium mass fraction $(X_{2P})$ as a function
of $\Gamma _3$ and, these bounds are used to constrain the present
baryon-to-photon ratio $(\eta)$ and baryon density $(\Omega_B)$.  For
$\Gamma _3 \geq$ 1/4 it is found that (from D and \he
alone): $3.1 \leq \eta_{10} \leq 9.0$; $0.045 \leq \Omega_B h^2_{50}
\leq 0.133$ (where $H_0 = 50h_{50}$ km$\,$s$^{-1}$ Mpc$^{-1}$).

\vfil\eject
\noindent
{\bf Introduction}

The light nuclides D, $^3$He, $^4$He and $^7$Li are produced in
astrophysically interesting abundances during big bang nucleosynthesis
(BBN) and, hence, their primordial abundances provide quantitative
tests of the standard hot big bang model.  However, primordial
abundances are not observed but, rather, derived from observational
data and -- to a greater or lesser extent -- models for galactic
chemical evolution.  Since $^4$He and $^7$Li are observed in metal-poor
objects (extragalactic HII regions for the former, PopII stars for the
latter), the evolutionary corrections are minimized.  Unfortunately,
this is not the case for D and \he where observations are, for the most
part, limited to the solar system and the present interstellar medium
(ISM).  The inevitable intervention of galactic evolution models
results in model dependent uncertainties for the primordial abundances
of D and \he derived from even the most accurate observational data.
Indeed, for this reason the most fruitful approach has been to use the
observational data to {\it bound} the primordial abundances.  For
example, since deuterium is only destroyed during the course of
galactic evolution, the primordial abundance must exceed the observed
abundance $(X_{2P} \geq X_{2{\rm OBS}}$ where $X_{2{\rm OBS}} =
X_{2\odot}$ or $X_{2{\rm ISM}})$.  It has, however, proved harder to
use the observational data to bound $X_{2P}$ from above (Yang et al.
1984, YTSSO).  YTSSO and Dearborn, Schramm \& Steigman (1986, DSS)
first noted that since D burns to \he and {\it some} \he survives
stellar evolution, the observed abundances of D and \he may be used to
bound, from above, the primordial abundance of D + \he. The upper bound
to the primordial abundance of D + \he derived by YTSSO, and utilized in
many subsequent analyses, depends on the \he survival fraction $g_3$
which DSS computed from models for the evolution of stars of different
masses.  However, unlike the model-independent {\it lower} bound
$(X_{2P} \geq X_{2{\rm OBS}})$, this upper bound to D + \he is still
model dependent in the sense that the ``effective'' value of $g _3$,
 hereafter indicated by $\Gamma _3$ to distinguish it from the DSS g$_3$,
depends not only on stellar structure/evolution but, also, on the past
star formation history, the IMF (and its evolution), infall and
outflow, etc.  In short, since the ``true'' value of $\Gamma _3$ is model
dependent, so too is the upper bound to the primordial abundance of
deuterium derived from observations of D and $^3$He.  Indeed, using very
much the same stellar data, Steigman \& Tosi (1992, ST92) find deuterium
destruction factors $\leq$ 2--3 while Vangioni-Flam \& Audouze (1988) and
Vangioni-Flam, Olive \& Prantzos (1994) find destruction factors as
large as $\sim 5-10$ are allowed and, the more extreme models of
Vangioni-Flam \& Cass\'e (1994) permit even more D destruction. We note
that these latter models have been designed to account for the D data and
they may not
necessarily be consistent with other observational constraints on the
abundances,
abundance ratios, and gradients of the other elements. Indeed, Edmunds
(1994) argues on
the \break \vfil\eject
\noindent basis of ``Simple Model" predictions, that they may not be
consistent with
the age-metallicity relation and the G-dwarf metallicity distribution.

Despite the above-cited model dependent uncertainties, it is important
to pursue the primordial abundance of deuterium.  The main reason is
that, in the context of BBN, $X_{2P}$ (the primordial deuterium mass
fraction) is a sensitive ``baryometer''.  At higher nucleon (baryon)
densities D burns faster (to $^3$H, $^3$He, $^4$He) and less deuterium
emerges from BBN.  $X_{2P}$ is a steeply decreasing function of the
nucleon-to-photon ratio $\eta [\eta_{10} = 10^{10}(N/\gamma)]$; for $2
\leq \eta_{10} \leq 6$, $X_{2P} \sim \eta^{-1.7}_{10}$.  Thus, an upper
(lower) bound to $X_{2P}$ will lead to a lower (upper) bound to $\eta$
and, hence, to bounds on the current universal density of baryons
$(\Omega_B = \rho_B/\rho_{\rm crit} = 0.015 \eta_{10} h^{-2}_{50}$
where $H_0 = 50 h_{50}$ km s$^{-1}$ Mpc$^{-1}$).  Furthermore, a lower
bound to $\eta$ (derived from an upper bound to $X_{2P}$), when combined
with an upper bound to the primordial $^4$He mass fraction, can be used
to bound the ``effective'' number of equivalent light neutrino degrees
of freedom (Steigman, Schramm \& Gunn 1977) and, hence, to constrain
``new'' physics (beyond the standard model of the strong and
electroweak interactions).  Thus, rather than ignore primordial
deuterium as a potential baryometer, it is surely more worthwhile to
try to understand the nature and extent of the model dependent
uncertainties in its inferred primordial abundance.  To this end, it is the
goal
here to derive simple analytical relations between the primordial and
observed abundances of D and \he which are general in the sense that
they do not rely on the special details of any specific chemical
evolution model.  The equations to be derived in the next section are
``generic'' but, they depend on one model dependent parameter, $\Gamma _3$,
the ``effective survival fraction of $^3$He''.  Different galactic
evolution models will yield different values of $\Gamma _3$ (as a function
of
time and of location in the Galaxy).  Highlighting the physical reasons
for the various values of $\Gamma _3$ may aid in identifying a
``reasonable''
range for $\Gamma _3$ and, hence, lead to nearly model independent bounds on
$X_{2P}$, $\eta_{10}$ and $\Omega_B$.

\bigskip
\ni
{\bf Evolution of Deuterium and Helium-3}

At any time $(t)$ consider a ``representative'' volume $(V)$ of the
ISM which, initially, contained an amount of gas $M_i$.  In general,
the gas mass in $V$ will change with time due to infall, outflow and
the sequestering (even if temporary) of gas in stars.  At any time
during the evolution of the Galaxy there will be a fraction, $f(t)$, of
the gas in the ISM (in $V$) which has never been through stars.  The
remaining fraction, $1-f(t)$, has been cycled through stars -- one or
more times -- and returned to the gas.  Thus, at time $t$ in volume $V$,
$f(t) M(t)$ has never been through stars and $(1-f(t)) M(t)$ has been
through one or more generations of stars.  Further, of the amount of
gas in $V$ at $t$, $M(t)$, there were initially (prior to any
evolution) $N_{2i}$ deuterium nuclei where,
$$N_{2i} = (X_{2i}/2) M_i/M_N\, . \eqno (1)$$
In (1), $X_{2i}$ is the initial (primordial) D mass fraction and $M_N$
is the nucleon mass.  Since any D cycled through stars is destroyed, at
time $t$ in volume $V$ only $N_2(t) \ D$-nuclei remain.
$$N_2(t) = f(t) (X_{2i}/2)(M(t))/M_N\, .\eqno (2)$$
Since $X_2(t) = 2 N_2(t) M_N/M(t)$, the D-survival factor $(f_2)$
and the virgin gas fraction $(f)$ are identical.
$$f_2(t) \equiv X_2(t)/X_{2i} = f(t)\, .\eqno (3)$$
Since deuterium is only destroyed, $f_2(t) \leq 1$.

Next, consider the evolution of $^3$He. \he is destroyed in the hot
interiors of all stars but preserved in the cooler outer layers (Iben
1967; Rood 1972).  For cooler low mass stars ($\leq 2.5$M$_\odot$)
hydrogen burning results in the net production of \he (Iben 1967; Rood
1972; Rood, Steigman \& Tinsley 1976; DSS).  Although net stellar
production of \he is potentially large, especially in the epoch between
the birth of the solar system and the present,
when the most effective $^3$He producers start to contribute to the ISM
enrichment,
it is also very
uncertain (Rood, Steigman \& Tinsley 1976; DSS;
Galli et al. 1994, Hogan 1994, Tosi et al. 1994).
Thus, in the spirit
of deriving ``conservative'' upper bounds to the primordial abundances
of D and $^3$He, newly synthesized \he (as distinct from prestellar D
burned to $^3$He) is ignored in the following.  This will lead to a {\it
lower} bound to $X_3(t)$, the \he mass fraction at any time $t$.
The quantitative effect of this neglect is reduced if presolar nebular
abundances are used rather than current interstellar values.  Now,
since a fraction $f(t)$ of the \he nuclei which were originally in
$M(t)$ have never been in stars and a fraction $1-f(t)$ have been
cycled through stars, for $X_3(t) = 3N_3(t) M_N/M(t)$ it follows that,
$$X_3(t) \geq X_{3i}\, f(t) + (1-f(t)) \Gamma _3(t) \left[X_{3i} +
3X_{2i}/2\right]\, .\eqno (4)$$
In (4), $\Gamma _3(t)$ is the ``effective survival fraction'' of $^3$He.
$\Gamma _3$
is model and time dependent since it depends not only on the stellar
models but, on the IMF and on the history of star formation (e.g.,
which stars have returned their material to the ISM by time $t$).  In
addition, although in material going through stars for the first time
the prestellar \he abundance is the sum of the D and \he primordial
abundances
(D is immediately burned to $^3$He), any material going through stars
for a second (or more) time will have had its initial deuterium already
destroyed; this, too, modifies $\Gamma _3$ from the simple DSS values $g_3$
and,
thus, $\Gamma _3$ depends on how much of the gas that has been through stars
at least once has been through subsequent generations.

The \he evolution factor $f_3(t)$ is,
$$f_3(t) \equiv X_3(t)/X_{3i} \geq f(t) + (1-f(t)) \Gamma _3(t)
(y_{23i}/y_{3i}) .\eqno (5)$$
In (5), $y_{23i}/y_{3i}$ is the initial ratio (by number) of $D +$ \he to
$^3$He. By combining equations (3) and (5) we may eliminate $f(t)$ and
directly relate $X_{2i}$ and $X_{3i}$ to $X_2(t)$ and $X_3(t)$.
However, before doing so, it is instructive to explore those conditions
under which $X_3$ will {\it increase} from its initial value $(f_3 >
1)$.  Eq. (5) may be rewritten as,
$$f_3(t) \geq 1+(1-f(t)) \left[\Gamma _3(t) (y_{23i}/y_{3i})-1\right]\,
.\eqno (6)$$
Since $1-f(t) \geq 0$, the \he mass fraction will increase if $\Gamma _3(t)
>
y_{3i}/y_{23i}$. As an illustration, if $\Gamma _3 \geq 1/4$ (DSS),
$(y_{23}/y_3)_P \geq 4$ is required; for standard BBN (Walker et al.
1991, WSSOK) this is satisfied for $\eta_{10} \leq 4.5$. Thus, if
$\Gamma _{3\odot} \geq 1/4$ and $\eta_{10} \leq 4.5$, it is predicted that
$X_{3\odot} \geq X_{3P} (\geq 2.9 \times 10^{-5})$; this is (as will be
seen) consistent with solar system data (Geiss 1993).  Alternatively
the ST92 models$^1$ suggest that $ \Gamma _{3\odot} \geq
1/2$ which implies $X_{3\odot} \geq X_{3P}$ for all $\eta_{10} \leq 10$.

\footnote {}
{\noindent
$^1$Due to unfortunate confusion in the DSS definition of   $g_3$ (which is
by mass and
not by number),
the survival fractions $f_3$ published by ST92 were underestimated on
average
by a factor 1.2. This implies $\Gamma _{3\odot} \approx \Gamma _{3ISM}
\approx 0.60$}

The model dependent factor $f_2(t) = f(t)$ may be eliminated from (5)
by the use of (3), relating the abundances (by number with respect to
hydrogen) of D and \he initially to those at any later time $t$.

$$y_{2i} \leq \left({X\over X_i}\right) \left\{\left[y_{23} +
\left({1\over \Gamma _3} - 1\right) y_3 \right] - {1\over \Gamma _3}
\left({y_{3i}\over y_{23i}}\right) y_{23}\right\}\, .\eqno (7)$$
In (7), $X = X(t)$ is the hydrogen mass fraction and $X_i$ is its
initial value.  The inequality is due to the neglect of stellar
synthesis of \he and is generic to all models of galactic evolution.
The only model dependence in (7) is in $\Gamma _3(t)$.  It will be noticed
that eq. (7) is a ``mixed'' expression in the sense that the last term
on the right hand side depends on {\it both} initial and final
abundances.  One approach (YTSSO) is to neglect this term,
strengthening the inequality; this is equivalent to assuming zero
initial $^3$He.  For $y_{3i} > 0$,
$$y_{2i} < y^0_{2i} \leq \left({X\over X_i}\right) \left[y_{23} +
\left({1\over \Gamma _3} - 1\right) y_3\right]\, .\eqno (8)$$
Except for the prefactor $(X/X_i)$, eq. (8) (for $y^0_{2i}$) is
precisely the expression derived in YTSSO and used in many subsequent
analyses.  However, this prefactor is required to account properly for
the changing hydrogen mass fraction.  Since $X(t) < X_i$ the ``usual''
bound to $y_{2i}$ is somewhat overestimated; for example, for
$X_P = 0.76 \pm 0.02$ and $X(t) = 0.70 \pm 0.02$, $X(t)/X_P = 0.92 \pm
0.04$.

The symbiotic relationship between D and \he is clear in equations (7)
and (8) where, for fixed (observed) abundances $y_2$ and $y_3$, the
initial abundances of D and \he \ are anticorrelated (e.g., higher
$y_{3i}$ forces lower $y_{2i}$).  Although eq. (8) provides a valuable
BBN-independent bound to pre-galactic deuterium (Steigman 1994), the
last term on the right hand side of (7) should not, in general, be
neglected.  This poses no problem in practice since, for a specific BBN
model the left and right hand sides of (7) may be compared to find if
the inequality is, or is not, satisfied.  In this manner the
observational data can be used, along with evolution model-dependent
choices of $\Gamma _3$, to constrain the models/parameters of BBN.

In subsequent comparisons, the solar system data of Geiss
(1993) will be adopted.  The observed quantities, derived from solar
wind, lunar and meteoritic data are the ratios (by number) to $^4$He of
presolar \hed and \he ($y_{23}/y_4$ and $y_3/y_4$).  With such data it
is convenient to rewrite (7) and (8) in the following forms,
$$X_{2i} \leq {Y_\odot\over 2} \left\{\left[1- {1\over \Gamma _{3\odot}}
\left({y_{3i}\over y_{23i}}\right)\right] \left({y_{23}\over
y_4}\right)_\odot + \left({1\over \Gamma _{3\odot}} -
1\right)\left({y_3\over
y_4}\right)_\odot \right\} , \eqno (9)$$
$$X^0_{2i} \leq {Y_\odot\over 2} \left[\left({y_{23}\over
y_4}\right)_\odot + \left({1\over \Gamma _{3\odot}} -
1\right)\left({y_3\over
y_4}\right)_\odot \right] .\eqno (10)$$
In (9) and (10), $Y_\odot$ is the solar $^4$He mass fraction.

\bigskip
\ni
{\bf Solar System Abundances}

In a very valuable recent review, Geiss (1993) has reanalysed the solar
system data crucial to determining the presolar nebular abundances of D
and $^3$He.  Although Geiss' central values are virtually identical to
those used in the past (e.g., Boesgaard \& Steigman 1985), his error
bars are more generous.  Here, with a little care towards error
propagation, the Geiss (1994) analysis is adopted.
$$\left(y_{23}/y_4\right)_\odot = 4.09 \pm 0.83 \times 10 ^{-4} ,\eqno
(11a)$$
$$\left(y_3/y_4\right)_\odot = 1.52 \pm 0.30 \times 10 ^{-4} . \eqno
(11b)$$
The adopted value of the solar $^4$He abundance, with generous
allowance for uncertainty, is
$$Y_\odot = 0.28 \pm 0.02 \ \ \left(y_{4\odot}= 0.10\pm 0.01\right) . \eqno
(12)$$
For $Z_\odot \approx 0.02$, $X_\odot = 0.70 \pm 0.02$ and the above
values may be used to derive:

$$\eqalignno{(Y_\odot/2)(y_{23}/y_4)_\odot &= 5.73\pm 1.23 \times
10^{-5} ,&(13a)\cr
(Y_\odot/2)(y_3/y_4)_\odot &= 2.13 \pm 0.45 \times 10^{-5} ,&(13b)\cr
X_{2\odot} &= 3.60 \pm 1.26 \times 10^{-5} , &(13c)\cr
X_{3\odot} &= 3.19 \pm 0.67 \times 10^{-5} , &(13d)\cr
y_{23\odot} &= 4.09 \pm 0.92 \times 10^{-5} ,& (13e)\cr
y_{3\odot} &= 1.52 \pm 0.34 \times 10^{-5} , &(13f)\cr
y_{2\odot} &= 2.57 \pm 0.92 \times 10^{-5} . &(13g)\cr}$$

\bigskip
\ni
{\bf Bounds To Primordial Deuterium}

Armed with solar system abundances, equations (3) and (9) (or (10)) may
be enlisted to provide upper and lower bounds to the pregalactic
(primordial) abundance of deuterium.  If equation (9) is used,
$y_{3i}/y_{23i}$ will be taken from standard BBN (e.g., WSSOK).  As already
emphasized,  the only galactic evolution model dependence is
contained in $\Gamma _{3\odot}$, the effective \he  survival fraction in the
solar neighbourhood at the time of the formation of the solar system.
For quantitative comparisons $\Gamma _{3\odot} = 1/4$   and $\Gamma
_{3\odot} =$
1/2  will be adopted; in general, the larger
$\Gamma _{3\odot}$, the smaller the upper bound to $X_{2P}$.

Since, at the time of formation of the solar system, at least some of
the material then present in the local ISM had been cycled through
stars, $f_\odot < 1$.  Thus, $X_{2P} > X_{2\odot}$; this differs from
the often quoted constraint $y_{2P} > y_{2\odot}$ by the factor
$X_\odot/X_P \approx 0.9$.  At the 2-sigma ($\sim 95$\% confidence)
level, it follows from (13c) that,
$$X_{2P} > X_{2\odot} \geq 1.08 \times 10^{-5} .\eqno (14)$$
Because of the relatively large uncertainty in $X_{2\odot}$, this only
provides a very weak constraint on $X_{2P}$.  For a more restrictive
lower bound to primordial deuterium, the ISM D/H ratio (Linsky et al.
1993) is more useful.  The interstellar observations from Copernicus
and IUE (see, e.g., Boesgaard \& Steigman 1985) and HST (Linsky et al.
1993) extend over two orders of magnitude in H column density and are
consistent with
$$y_{2{\rm ISM}} = 1.6 \pm 0.2 \times 10^{-5} . \eqno  (15)$$
If $X_{\rm ISM} \approx X_\odot \approx 0.70 \pm 0.02$, then at
2-sigma,
$$X_{2P} > X_{2{\rm ISM}} > 1.7 \times 10^{-5} . \eqno (16)$$
For standard BBN (WSSOK) this provides an upper bound to the nucleon
density,
$$\eta_{10} < 9.0 , \eqno (17)$$
where $\eta$ is the ratio of nucleons to photons at present and
$\eta_{10} \equiv 10^{10}\eta$.

As an aside, notice that if the solar system (13g) and interstellar
(15) D/H ratios are compared,
$$X_{2\odot}/X_{2{\rm ISM}} \approx y_{2\odot}/y_{2{\rm ISM}} = 1.6 \pm
0.6 . \eqno (18)$$
This provides a weak ($\sim 1$-sigma) hint that, in the solar
neighbourhood, some D has been destroyed in the last 4.6 Gyr.

In considering the upper bound to the primordial deuterium mass
fraction from equations (9) or (10), it is useful to compare the BBN
predicted value $X_{2{\rm BBN}}$ (as a function of $\eta$) with the
analytic bounds $X_{2{\rm MAX}}$ (eq. 9) and $X^0_{2{\rm MAX}}$ (eq.
10).  For a ``2$\sigma$'' upper bound, equations (9) and (10) are
evaluated with the 2$\sigma$ upper bounds to the solar system
abundances (see 13a \& b: $(Y_\odot/2)(y_{23}/y_4)_\odot \leq 8.2 \times
10^{-5}$, $(Y_\odot/2)(y_3/y_4)_\odot \leq 3.0 \times 10^{-5}$). The
results for $X_{2{\rm MAX}}$ and $X^0_{2{\rm MAX}}$
are shown in Figure 1 for $\Gamma _3 = 1/4\ \& \ 1/2$ respectively.  Also
shown in Figure 1 is $X_{2{\rm BBN}}$; the requirement that $X_{2{\rm
BBN}} < X_{2{\rm MAX}}$ (or $X^0_{2{\rm MAX}}$) provides a lower bound
to $\eta$.

As anticipated, the neglect of any primordial \he $\left(X^0_{2{\rm
MAX}}\right)$
provides a weaker constraint than if the BBN predicted relative $^3$He/D
abundance is allowed for ($X_{2{\rm MAX}}$); for all $y_{3P} > 0$, \xet
$>$ \xe.  As $\eta$ increases, the BBN yields of D and \he decrease
but, ($^3$He/D)$_{\rm BBN}$ increases.  It is the increasing importance of
\he at high $\eta$ which is responsible for the decrease in \xe.

As more \he survives stellar processing ($\Gamma _3$ increases), consistency
with the solar system abundances restricts primordial D to lower values
(\xe and \xet decrease).  Thus, for fixed observed abundances, the
lower $\Gamma _{3\odot}$, the weaker the constraint on $X_{2P}$.  For
$\Gamma _{3\odot}
\geq 1/4$, $X_{2P} <$ \xet $\leq 16 \times 10
^{-5}$ (@ 2$\sigma$) which, for BBN, corresponds to $\eta^0_{10} \geq
2.5$. However, for $\eta_{10} \geq 2.5$, $y_{3P}/y_{23P} \geq 0.14$ and
the neglect of primordial \he is non-negligible.  From Figure 1 it may
be seen that for $\Gamma _{3\odot} \geq 1/4$, $X_{2{\rm BBN}} <$ \xe
(@2$\sigma$) for $X_{2P} \leq 11 \times 10^{-5}$; this corresponds to
$\eta^{\rm MIN}_{10} > 3.1$. For the same solar system data, if
$\Gamma _{3\odot}$ increases to $\simgt 1/2$, $\eta^{\rm MIN}_{10}$
increases
to $\simgt 4.0$.

In Figure 2, $X_{2{\rm BBN}}$ is compared to the $2\sigma$ {\it upper}
bound to \xe for $\Gamma _{3\odot} = 1/4$ and, to the $2\sigma$ {\it lower}
bound
to $X_{2{\rm MIN}}$ from the interstellar D/H ratio (eq. 16).  From D
and \he alone, $\eta$ is then restricted to the range $3.1 \leq \eta_{10}
\leq 9.0$.  Observational data on $^4$He and $^7$Li (e.g. WSSOK; Olive
\& Steigman 1994) further constrain the upper bound to $\eta (\eta_{10}
\leq 3.9)$.

Earlier, it had been noted that, provided $\Gamma _3$ is not too small nor
$y_{3P}/y_{23P}$ too large, it is likely that the \he mass fraction
will increase from its primordial value (even neglecting net stellar
production of $^3$He).  Thus, for $\Gamma _{3\odot} \geq 1/4$ and $\eta_{10}
\leq 4.5$, it was expected that $X_{3\odot} \geq X_{3{\rm BBN}}$.  This
upper bound to primordial \he will also lead to a lower bound to $\eta$
which, although weaker than that above, is still of some interest.
 From (13d), @$2\sigma$, $X_{3P} \leq X_{3\odot} \leq 4.5 \times 10 ^{-3}$,
which
corresponds to $\eta_{10} \geq 2.0$. Since this bound is independent of the
previous
deuterium constraint it may be used to bound primordial/pregalactic D: for
$\eta_{10}
\geq 2.0$, $(D/H)_P \leq 1.6 \times 10^{-4}$ (and, for the standard model of
3 light neutrino species, $Y_P \geq 0.236)$.  This bound is of some
interest considering the claim of a possible detection of deuterium in a QSO
absorption system at a level of $(D/H)_{\rm QSO} \approx 1.9-2.5 \times
10^{-4}$ (Songaila et al. 1994; Carswell et al. 1994).

\bigskip
\ni
{\bf Comparison With Previous Bounds}

The bounds of primordial deuterium derived above are quantitatively
similar to previous bounds (YTSSO; WSSOK; ST92) but, there are some
qualitative differences.  One source of these differences is the new
Geiss (1993) solar system abundances; another is the use of mass
fractions $(X_2)$ rather than number ratios to hydrogen ($y_2)$.  For
example, YTSSO had derived a version of equation (8) which neglected
$X_\odot/X_P (\approx 0.92)$ and related $y_{23P}$ (rather than
$y_{2P}$; $y_{2P} < y_{23P})$ to $y_{23\odot}$ and $y_{3\odot}$: $y_{23P}
< y_{23\odot} + (g^{-1}_{3\odot} - 1) y_{3\odot}$.  For the $2\sigma$
upper bounds to \he and \hed, YTSSO took: $y_{23\odot} \leq 4.3
\times 10^{-5}$, $y_{3\odot} \leq 1.9 \times 10^{-5}$, which leads to:
$10^5 y_{23P} < 2.4 + 1.9 g^{-1}_{3\odot}$. Then, for $g_{3\odot} \geq
1/4$, YTSSO derive $y_{23P} \leq 10 \times 10^{-5}$ which, corresponds
to $\eta_{10} \geq 3$ (YTSSO); actually, to two significant figures,
$\eta_{10} \geq 2.8$. WSSOK used the YTSSO constraint but, slightly
different $2\sigma$ bounds to solar system \hed and \he to derive:
$10^5 y_{23P} < 3.1 + 1.8g^{-1}_{3\odot}$. For
$g _{3\odot} \geq 1/4$,
WSSOK find $y_{23P} < 10.3 \times 10^{-5}$ (WSSOK also derive the constraint
$y_{23P}X_P < 7.5 \times 10^{-5}$ which, for $Y_P \leq 0.25$, results in
$y_{23P} \leq 10 \times 10^{-5}$); this, too, corresponds to $\eta_{10}
\geq 2.8$ (WSSOK).  Finally, ST92 used the DSS stellar
models for $g_3$ in the standard galactic evolution models of Tosi
(1988) to identify the ranges of $X_{2P}$ and $X_{3P}$ allowed by the
solar system and ISM data.  For the $2\sigma$ upper bounds to solar
system abundances ST92 chose $X_{2\odot} \leq 4.5 \times 10^{-5}$ and
$X_{3\odot} \leq 3.6 \times 10^{-5}$; for ISM deuterium they adopted
$X_{2{\rm ISM}} \leq 4.8 \times 10^{-5}$.  In these models the
effective \he survival fraction is large ($\Gamma _{3\odot} \geq 1/2$, see
the
earlier footnote) and the BBN constraint more restrictive: $X_{2P} \leq 9.0
\times 10^{-5}$; $\eta_{10} \geq 3.7$. Note that if the ST92 adopted
abundances
are used in eq. (10), $10^5 X^0_{2P} \leq 4.2 + 2.4\Gamma^{-1}_{3\odot}$,
which, for $\Gamma _{3\odot} \geq 1/4$, implies $X^0_{2P} \leq 13.8 \times
10^{-5}$ and $\eta^0_{10} \simgt 2.7$; this is to be compared to
$X^0_{2P} \leq 16 \times 10^{-5}$ and $\eta^0_{10} \simgt 2.5$ using
the Geiss (1993) abundances.
Thus, although the Geiss abundances loosen the restrictions on
$\eta^0_{10}$ compared with YTSSO and WSSOK, it is seen that the
neglect of primordial \he is significant and, the Geiss abundances
in eq. (9) (with $\Gamma _{3\odot} \geq 1/4)$, lead to the more restrictive
constraint found above: $X^{\rm MAX}_{2P} \leq 11 \times 10^{-5}$;
$\eta^{\rm MIN}_{10} > 3.1$.

\bigskip
\ni
{\bf Summary and Implications}

The primordial abundances of D and \he cannot be derived from solar
system or ISM data (with the exception that $X_{2P} \geq X_{2\odot}$,
$X_{2P} \geq X_{2{\rm ISM}}$) without the intervention of a model to
follow the galactic evolution of these nuclides.  It has been the goal
here to derive a ``generic'' analytic expression relating primordial
and galactic abundances.  Equation (9) incorporates the effects (and
uncertainties!) of stellar and galactic evolution in one free
parameter $\Gamma _3$.  For $\Gamma _{3\odot} \geq 1/4$ (DSS) and the Geiss
(1993) solar
system abundances it has been found, neglecting net stellar
production of \he in low mass stars (Iben 1967, Rood 1972) that, $X_{2P} <
X^{\rm MAX}_{2P} \leq 11 \times 10^{-5}$.  For standard BBN (WSSOK),
this corresponds to $y_{2P} < 7.3 \times 10^{-5}$ and $\eta_{10} > 3.1$.
This provides a slightly more restrictive lower bound to the universal
abundance of nucleons (e.g. compare to WSSOK where $\eta_{10} \geq
2.8$). For standard ($N_\nu = 3$) BBN the lower bound to the predicted
primordial mass fraction of $^4$He increases from the WSSOK value to
$Y_{\rm BBN}
\geq 0.241 \pm 0.001$  as does the lower
bound to primordial lithium: (Li/H)$_{\rm BBN} \geq 1.2 \times
10^{-10}$. For $\eta_{10} \geq 3.1$, $\Omega_B \geq 0.045 h^{-2}_{50}$
so that, for $H_0 \leq 100$ km s$^{-1}$ Mpc$^{-1}$, $\Omega_B \geq
0.011$. This strengthens, very slightly, the case for baryonic dark
matter $(\Omega_B > \Omega_{\rm LUM}$, see WSSOK).

The upper bound to $\eta$ from the lower bound to primordial D
$(X_{2P} \geq 1.7 \times 10^{-5}$, $\eta_{10} \leq 9.0)$ is weaker than
that from lithium [(Li/H)$_P \leq 2 \times 10^{10}$, $\eta_{10} \leq
4.0$] or from $^4$He ($Y_P \leq 0.243$, $\eta_{10} \leq 3.9)$. If this
bound were saturated ($\eta_{10} \approx 9.0)$, the primordial lithium
abundance would be $\sim$half solar and $Y_P \approx 0.253 \pm 0.001$
(for $N_\nu = 3$).

In closing, it must be reemphasized that the analytic framework
developed here is not a replacement for detailed evolution models.  In
the absence of such models, the choice of the \he survival fraction,
$\Gamma _3$, is little more than a guess.  Rather, the potential value of
this framework is that it provides a concise way to compare the results
of different models; models which yield different constraints using the
same observational data must differ in $\Gamma _3$.  Tracing the source(s)
of
these differences may help to choose among different evolution models
and/or to decide on the true uncertainties in the derived primordial
abundances.

\bigskip
We thank Daniele Galli for calling our attention to the issue of the correct
meaning of the $g_3$ values published by DSS and adopted by ST92. This work
was begun
when G. S. was an Overseas Fellow at Churchill College, Cambridge and a
Visiting Fellow
at the Institute of Astronomy and he thanks them for hospitality. The work
at OSU is
supported by DOE grant DE-FG02-94ER-40823.

\vfil\eject
\ni
{\bf References}

\ref
Boesgaard, A.M. and Steigman, G., Ann. Rev. Astron. Astrophys.
{\bf 23}, 319 (1985).
\ref
Carswell, R.F., Rauch, M., Weymann, R.J., Cooke, A.J. and Webb, J.K.,
\MN {\bf 268}, L1 (1994).
\ref
Dearborn, D.S.P., Schramm, D.N. and Steigman, G., \ApJ {\bf 302}, 35
(1986). (DSS)
\ref
Edmunds, M., \MN {\bf 270}, L37 (1994).
\ref
Galli, D., Palla, F., Straniero, O., Ferrini, F., \ApJ {\bf 432}, L101
(1994,)
\ref
Geiss, J., in {\it Origin and Evolution of the Elements} (eds. N.
Prantzos, E. Vangioni-Flam and M. Cass\'e; Cambridge Univ. Press),
p.89 (1993).
\ref
Hogan, C.J., University of Washington preprint (1994)
\ref
Iben, I., \ApJ {\bf 147}, 624 (1967).
\ref
Linsky, J.L., Brown, A., Gayley, K., Diplas, A., Savage, B.D., Ayres,
T.R., Landsman, W., Shore, S.W. and Heap, S., \ApJ {\bf 402}, 694
(1993).
\ref
Olive, K.A. and Steigman, G., \ApJS (In Press). (OSU-TA-6/94)
\ref
Rood, R.T., \ApJ {\bf 177}, 681 (1972).
\ref
Rood, R.T., Steigman, G. and Tinsley, B.M., \ApJ (Lett) {\bf 207},
L57 (1976).
\ref
Songaila, A., Cowie, L.L., Hogan, C. and Rugers, M., Nature {\bf 368}, 599
(1994).
\ref
Steigman, G. and Tosi, M., \ApJ {\bf 401}, 150 (1992). (ST92)
\ref
Steigman, G., Schramm, D.N. and Gunn, J.E., Phys. Lett. {\bf B66},
202 (1977)
\ref
Steigman, G., \MN {\bf 269}, L53 (1994).
\ref
Steigman, G. and Walker, T.P., In Preparation (1994).
\ref
Tosi, M., A\&A {\bf 197}, 33 (1988).
\ref
Vangioni-Flam, E. and Audouze, J.,  A\&A {\bf 193}, 81 (1988).
\ref
Vangioni-Flam, E., Olive, K.A. and Prantzos, N., \ApJ {\bf 427}, 618 (1994).
\ref
Vangioni-Flam, E. and Cass\'e, M., Preprint (1994).
\ref
Walker, T.P., Steigman, G., Schramm, D.N., Olive, K.A. and Kang, H.S.
\ApJ {\bf 376}, 51 (1991). (WSSOK)
\ref
Yang, J., Turner, M.S., Steigman, G., Schramm, D.N. and Olive, K.A.
\ApJ {\bf 281}, 493 (1984). (YTSSO)
\newpage
\centreline{\bf Figure Captions}

\item {Figure 1a.} The primordial deuterium mass fraction $(X_{2P})$ as a
function of the present nucleon-to-photon ratio $\eta_{10}$.  The solid
curve $(X_{2{\rm BBN}}$) is the standard BBN prediction.  The dotted
curve ($X^{2\sigma,0}_{2{\rm MAX}}$) and the dashed curve
($X^{2\sigma}_{2{\rm MAX}}$) are computed from equations (10) and (9)
respectively using the $2\sigma$ upper bounds to the abundances in eq.
(13) for $\Gamma _{3\odot} = 1/4$.

\item {Figure 1b.} As in Figure 1a, for $\Gamma _{3\odot} = 1/2$.

\item {Figure 2.} Upper and lower bounds to primordial deuterium.  The
solid curve ($X_{2{\rm BBN}}$) and the dashed curve
($X^{2\sigma}_{2{\rm MAX}}$) are as in Figure 1. The dotted line
($X^{2\sigma}_{2{\rm MIN}}$) is the $2\sigma$ lower bound to $X_{2P}$
from the ISM data (eq. 16).  The restriction $X^{2\sigma}_{2{\rm MIN}}
\leq X_{2{\rm BBN}} \leq X^{2\sigma}_{2{\rm MAX}}$ bounds $\eta_{10} :
3.1 \leq \eta_{10} \leq 9.0$.

\end